\documentclass[aps,prb,twocolumn,floats,showpacs]{revtex4}
\usepackage{epsfig}
\usepackage{bm}
\usepackage{latexsym}

\begin{document}

\newcommand{\hide}[1]{}
\newcommand{\tbox}[1]{\mbox{\tiny #1}}
\newcommand{\half}{\mbox{\small $\frac{1}{2}$}}
\newcommand{\sinc}{\mbox{sinc}}
\newcommand{\const}{\mbox{const}}
\newcommand{\trc}{\mbox{trace}}
\newcommand{\intt}{\int\!\!\!\!\int }
\newcommand{\ointt}{\int\!\!\!\!\int\!\!\!\!\!\circ\ }
\newcommand{\eexp}{\mbox{e}^}
\newcommand{\bra}{\left\langle}
\newcommand{\ket}{\right\rangle}
\newcommand{\EPS} {\mbox{\LARGE $\epsilon$}}
\newcommand{\ar}{\mathsf r}
\newcommand{\im}{\mbox{Im}}
\newcommand{\re}{\mbox{Re}}
\newcommand{\bmsf}[1]{\bm{\mathsf{#1}}}


\title{Probing the eigenfunction fractality with a stop watch}
\author{J. A. M\'endez-Berm\'udez and Tsampikos Kottos
\\
Max-Planck-Institut f\"ur Dynamik und Selbstorganisation, Bunsenstra\ss e 10,
D-37073 G\"ottingen, Germany
}

\begin{abstract}
We study numerically the distribution of scattering phases ${\cal P}(\Phi)$ and of Wigner delay times ${\cal P}(\tau_W)$ for the power-law banded random matrix (PBRM) model at criticality with one channel attached to it. We find that ${\cal P}(\Phi)$ is insensitive to the position of the channel and undergoes a transition towards uniformity as the bandwidth $b$ of the PBRM model increases. The inverse moments of Wigner delay times scale as $\langle\tau_W^{-q} \rangle\sim L^{- q D_{q+1}}$, where $D_q$ are the multifractal dimensions of the eigenfunctions of the corresponding closed system and $L$ is the system size. The latter scaling law is sensitive to the position of the channel.
\end{abstract}
\maketitle

\section{Introduction}

The transport properties of random media have been a subject of intensive research activity
for more than fifty years~\cite{A58,AMPJ95,AKL91,E97,SSSLS93,M00,FE95,EM00,V03,W80,AS86,BHMM01,
HK99}. The most fascinating aspect of this study is the appearance (in high-dimensions) of
a metal-insulator phase transition (MIT) as an external parameter changes. In the metallic
phase, the eigenstates are extended \cite{AKL91,E97,M00,FE95} and the statistical properties
of the spectrum are described well by random matrix theory \cite{AMPJ95,AKL91,E97,SSSLS93}.
In the localized phase, the levels become uncorrelated leading to a Poissonian level spacing
distribution \cite{SSSLS93}, and the eigenfunctions are exponentially localized \cite{A58,
AMPJ95,AKL91,E97,M00}.

The MIT where the phase transition from localized to extended states occurs, is characterized
by remarkably rich critical properties \cite{A58,BHMM01,AS86,M00,FE95,EM00,V03,W80}. In particular,
the eigenfunctions show strong fluctuations on all length scales and represent multifractal
distributions \cite{M00,FE95,W80}. This is one of the most important features at MIT and has
considerably helped our understanding of different phenomena in mesoscopic systems \cite{AMPJ95}.
The multifractal structure of the eigenfunctions is usually quantified by studying the size
dependence of the so-called participation numbers (PN)
\begin{equation}
\label{PN}
{\cal N}_q = \left(\int \left|\psi({\bf r})\right|^{2q} d{\bf r}\right)^{-1}
\propto L^{(q-1)D_{q}}
\end{equation}
where $L$ is the linear size of the system and $D_q$ are the multifractal dimensions of the 
eigenfunction $\psi({\bf r})$. Among all the dimensions, the correlation dimension 
$D_2$ plays the most prominent role. The corresponding PN is roughly equal to the number of 
non-zero eigenfunction components, and therefore is a good and widely accepted measure of the 
extension of the states. At the same time, $D_2$ manifest itself in a variety of other physical 
observables. As examples we mention the conductance distribution in metals \cite{AKL91,BHMM01}, 
the statistical properties of the spectrum \cite{BHMM01}, the anomalous spreading of a wave-packet,
and the spatial dispersion of the diffusion coefficient \cite{HK99}.

It is therefore highly desirable to have the possibility of measuring the multifractal dimensions
$D_q$, and specifically $D_2$, directly from an experiment. Unfortunately, to our knowledge,
there is no experimental study of eigenfunction
fluctuations at the transition regime \cite{note2}, although in recent years a MIT was
experimentally accessible in microwave \cite{CSG00} and optical wave \cite{WBLR00} setups.
The main difficulty is the lack of eigenfunction accessibility in all these experiments. On
the other hand, new techniques were developed that allow accessibility to various quantities
associated with the scattering process. Among them is the Wigner delay time \cite{CG01,JPM03,
LGISS01} which captures the time-dependent aspects of quantum scattering. It can be interpreted
as the typical time an almost monochromatic wave packet remains in the interaction region.
It is related to the energy derivative of the total phase shift $\Phi(E)= -i\ln \det S(E)$
of the scattering matrix $S (E)$, i.e. $\tau_W (E) = {1\over M} {d\Phi(E) \over dE}$, where
$M$ is the number of channels.

These experimental efforts have promoted the study of Wigner delay times to quantities of
interest in their own right and attracted a lot of theoretical work. For chaotic systems
many results are known concerning the distribution of Wigner delay times ${\cal P}(\tau_W)$
\cite{FS97}. Recently, the interest has extended to systems showing diffusion \cite{OKG03,F03}
and localization \cite{TC99}. At the same time, an intensive activity to understand ${\cal P}
(\tau_W)$ for systems at critical conditions was undertaken in \cite{KW02,F03,MK04,OF04}. A
possibility of anomalous scaling of inverse moments of $\tau_W$ with the system size $L$
was suggested in \cite{F03}. Specific predictions linking the scaling exponents and the
multifractal properties of eigenfunctions of the corresponding closed system were made in
\cite{MK04,OF04}. While in both cases an anomalous scaling of the inverse moments of delay
times was reported, there was a discrepancy in the exact power dictating these scaling laws.

Precise reasons for these discrepancies are still not clear (see though the discussion in
\cite{OF04}) and deserve investigation. Here we undertake this task by studying numerically
the distribution of Wigner delay times at MIT for the prototype power-law banded random matrix
(PBRM) model with one channel attached to it. We provide clear evidence showing that the inverse
moments of Wigner delay times $\langle \tau_W^{-q} \rangle$ scale as
\begin{equation}
\label{MIT}
\langle \tau_W^{-q} \rangle \propto L^{-f(q)},\quad f(q)= q D_{q+1}
\end{equation}
where $\langle . \rangle$ stands for an ensemble average in agreement with the theoretical
prediction of \cite{OF04}. However we find that this relation is extremely fragile. Namely
we show that it holds for channels attached to a {\it typical} position in the sample. At the
same time we perform a detail analysis of the distribution of phases ${\cal P}(\Phi)$ of the
scattering matrix associated with our scattering setup. We show that ${\cal P}(\Phi)$ is insensitive
to the position of the channel while it depends drastically on the band-width $b$ which characterize
the criticality. It undergoes a transition towards uniformity as $b$ increases.
We note that the theoretical analysis of \cite{OF04} applies only for $b\gg 1$ while our numerical investigations extends over the whole $b$-range.

The structure of this paper is as follows. In the next section we describe the scattering setup
and summarize the properties of the corresponding closed system. Section III discuss the distribution of phases for various bandwidths, coupling strengths, and position of the channel. In Section IV we analyze the statistical properties of delay times. Finally, our conclusions are given in Section V.

\section{Scattering setup}

The isolated sample is represented by $L\times L$ real symmetric matrices whose entries are
randomly drawn from a normal distribution with zero mean and a variance
depending on the distance of the matrix element from the diagonal
\begin{equation}
\label{pbrm}
\left<(H_{ij})^2\right> = {1\over 1+(|i-j|/b)^{2\alpha}}\, \, ,
\end{equation}
where $b$ and $\alpha$ are parameters. In a straightforward interpretation, the PBRM model describes
a $1$D sample with random long-range hopping. Also, the PBRM ensemble arises as an effective
description to a variety of physical systems, such as the quantum Fermi accelerator, the
scattering by a Coulomb center in an integrable billiard, and in billiard systems with
non-analytic boundaries.

Field-theoretical considerations \cite{MFDQS96,M00,KT00} and detail numerical investigations
\cite{EM00,V03} verified that the model undergo a transition at $\alpha=1$ from localized
states for $\alpha >1$ to delocalized states for $\alpha < 1$. This transition shows all the
key features of the Anderson MIT, including multifractality of eigenfunctions and non-trivial
spectral statistics at the critical point. At the center of the spectral-band a theoretical
estimation for the multifractal dimensions $D_q$ gives \cite{MFDQS96}
\begin{equation}
\label{Dq}
D_q =\left\{
\begin{array}{cc}
 4b \Gamma(q-1/2)[\sqrt{\pi}(q-1)\Gamma(q-1)]^{-1} \ &, b \ll 1 \nonumber\\
 1-q(2\pi b)^{-1} \   &, b\gg 1
\end{array}
\right.
\end{equation}
where $\Gamma$ is the Gamma function. Thus model (\ref{pbrm}) possesses a line of critical
points $b\in (0,\infty)$, where the multifractal dimensions $D_q$ change with $b$.

We turn the isolated system to a scattering one by attaching one semi-infinite single channel
lead to it. The lead is described by a 1D semi-infinite tight-binding hamiltonian
\begin{equation}
\label{leads}
H_{\rm lead}=\sum_{n=1}^{-\infty} (|n><n+1| + |n+1><n|)\,\,.
\end{equation}

Using standard methods \cite{MW69} one can write the scattering matrix in the form \cite{KW02}
\begin{equation}
\label{smatrix}
S(E) = e^{i\Phi(E)}= {\bf 1}-2i \pi \, W^{\,T} {1\over E-{\cal H}_{\rm eff}}W,
\end{equation}
where ${\cal H}_{\rm eff}$ is an effective non-hermitian Hamiltonian given by
\begin{equation}
\label{Heff}
{\mathcal{H}}_{\rm eff}=H- i \pi WW^{\,T}.
\end{equation}
Here $W_{nm}=w_0 \delta_{nn_0}$ is a $L\times 1$ vector where $w_0$ is the coupling strength
and $n_0$ is the site at which we attach the lead. All our calculations take place in an
energy window close to the band center $(E=0)$. The delay time is given by \cite{FS97,KW02}
\begin{equation}
\label{tauW}
\tau_W(E=0) = \left. {d\Phi(E)\over dE} \right|_{E=0} = \left. -2 {\cal I}m {\rm Tr}(E-{\cal H}_{\rm eff})^{-1}\right|_{E=0}
\end{equation}

We will see latter on that the scaling
properties of inverse moments of delay times are independent of $w_0$ while they depend
drastically on the position $n_0$. Specifically, if $n_0$ is a representative site (like any
site in the bulk of the sample) where ergodicity of wavefunctions can be assumed, then
Eq~(\ref{MIT}) applies. Instead if $n_0$ belongs to the boundary (i.e. $n_0=1$ or $L$),
then Eq~(\ref{MIT}) does not valid any more.

In our numerical experiments bellow we used matrices of size varying from $L=50$ up to $L=2000$.
For better statistics a considerable number of different disorder realizations has been used. In
all cases we had at least $100000$ data for statistical processing.

\section{Distribution of phases}

We start our analysis with the distribution ${\cal P}(\Phi)$ of phases of the $S-$matrix.
Analytical considerations applied to the large $b\gg 1$ limit \cite{OF04} led to the following
expression
\begin{equation}
\label{phase}
{\cal P}(\Phi) = {1\over 2\pi (\gamma + \sqrt{\gamma^2-1} \cos(\Phi))}
\end{equation}
where $\gamma = (1+|\langle S\rangle|^2)/(1-|\langle S\rangle|^2)$. Notice that in the
limit of $|\langle S\rangle|=0$ we recover the uniform distribution, where the corresponding
setup is associated with perfect coupling of the sample to the leads. Eq.~(\ref{phase})
was also derived in the framework of Random Matrix Theory \cite{M94} using the maximum
entropy ansatz. It is interesting that the same expression applies (see bellow) in our
case as well.

\begin{figure}
\begin{center}
    \epsfxsize=8.4cm
    \leavevmode
    \epsffile{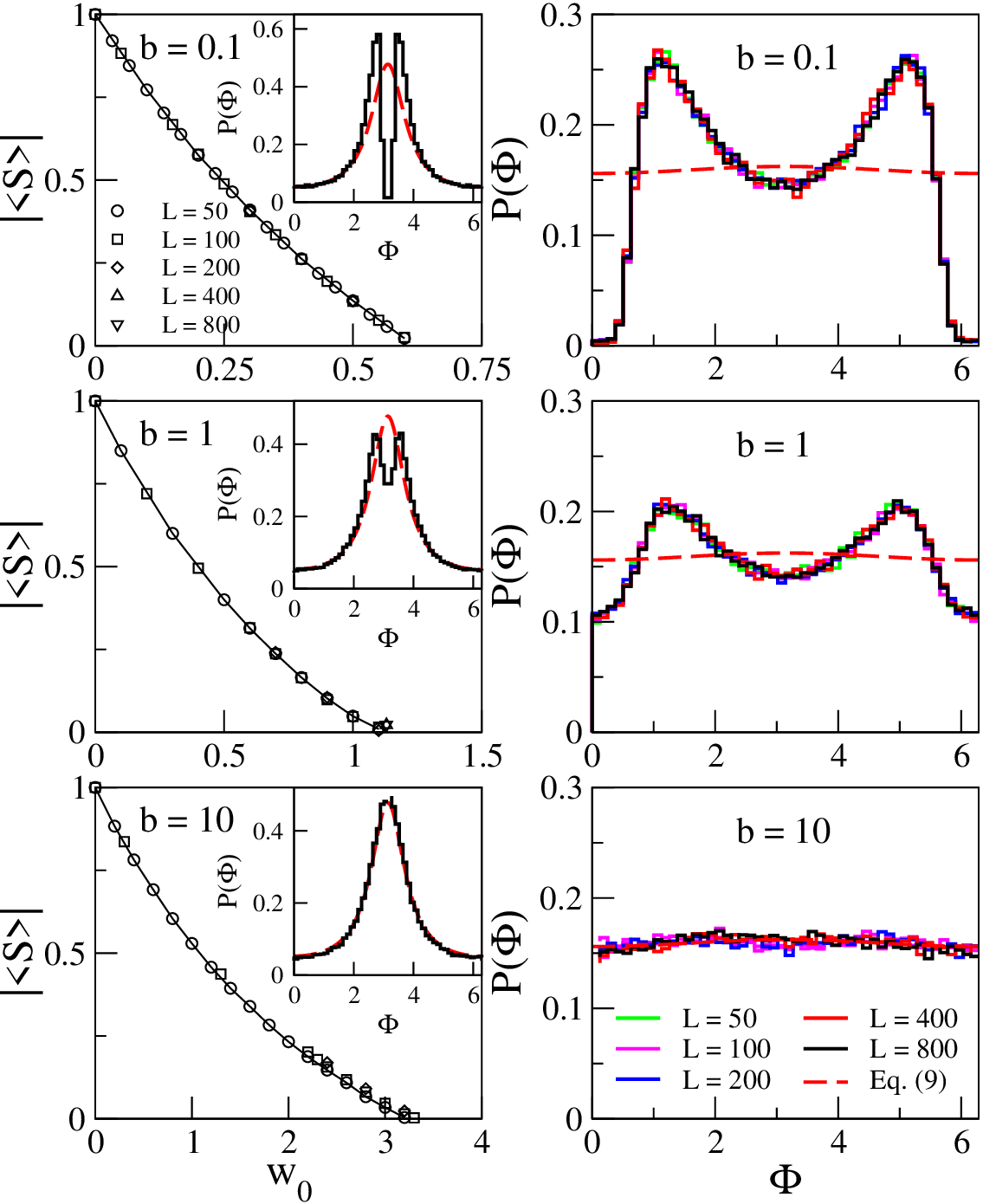}
\caption{Left panels: $|\langle S\rangle|$ as a function of $w_0$ for a channel attached to
the first site (boundary) of the sample. Right panels: ${\cal P}(\Phi)$ corresponding to $w_0$
where $|\langle S\rangle|$ takes its minimum value (see left panels) and for various system sizes,
ranging from $L=50$ up to $L=800$. In the insets of the left panels we show a representative
${\cal P}(\Phi)$ for $|\langle S\rangle|\sim 0.5$. The dashed line is
Eq.~(\ref{phase}). From top to bottom: $b=0.1$, 1, and 10.}
\label{fig:fig1}
\end{center}
\end{figure}

In Fig.~\ref{fig:fig1} we report our data  for various values of $b$ and various coupling
strengths $w_0$. The scattering setup corresponds to one channel attached to the boundary
of the sample. For large $b$ values, ${\cal P}(\Phi)$ agrees quite good with the theoretical
prediction of Eq.~(\ref{phase}) while for $b\leq 1$ we observe large deviations for any
$w_0$, even for $|\langle S(w_0)\rangle|\approx 0$.
For $|\langle S(w_0)\rangle| \neq 0$ and small values of $b$, two peaks appear around
$\Phi=\pi$ while a gap is created between them. As $b$ increases further, the two peaks
move closer to one another forming one central peak at $\Phi=\pi$. In the limit of
$|\langle S(w_0)\rangle|\rightarrow 1$ the distribution becomes a $\delta-$function
signifying a total reflection at the point where the lead is attached. Similar
results (not reported here) are found for the case where the lead is attached to the
bulk of the sample.

\section{Distribution of delay times}

In this section we analyze the distribution of Wigner delay times (\ref{tauW}). Recent investigations
\cite{MK04,OF04} suggested that inverse moments of $\tau_W$ scale in an anomalous way with respect
to the sample size $L$. It was further suggested that the scaling exponent is associated with the
multifractal dimensions of the eigenfunctions of the corresponding closed system. This approach
allow for an experimental measurement of the fractality of the wavefunctions with scattering techniques.
However, these contributions concluded on a different scaling exponent. It is therefore desirable
to investigate the origin of the discrepancy and specify the conditions under which the theoretical
expression (\ref{MIT}) is applicable.

\begin{figure}
\begin{center}
    \epsfxsize=8.4cm
    \leavevmode
    \epsffile{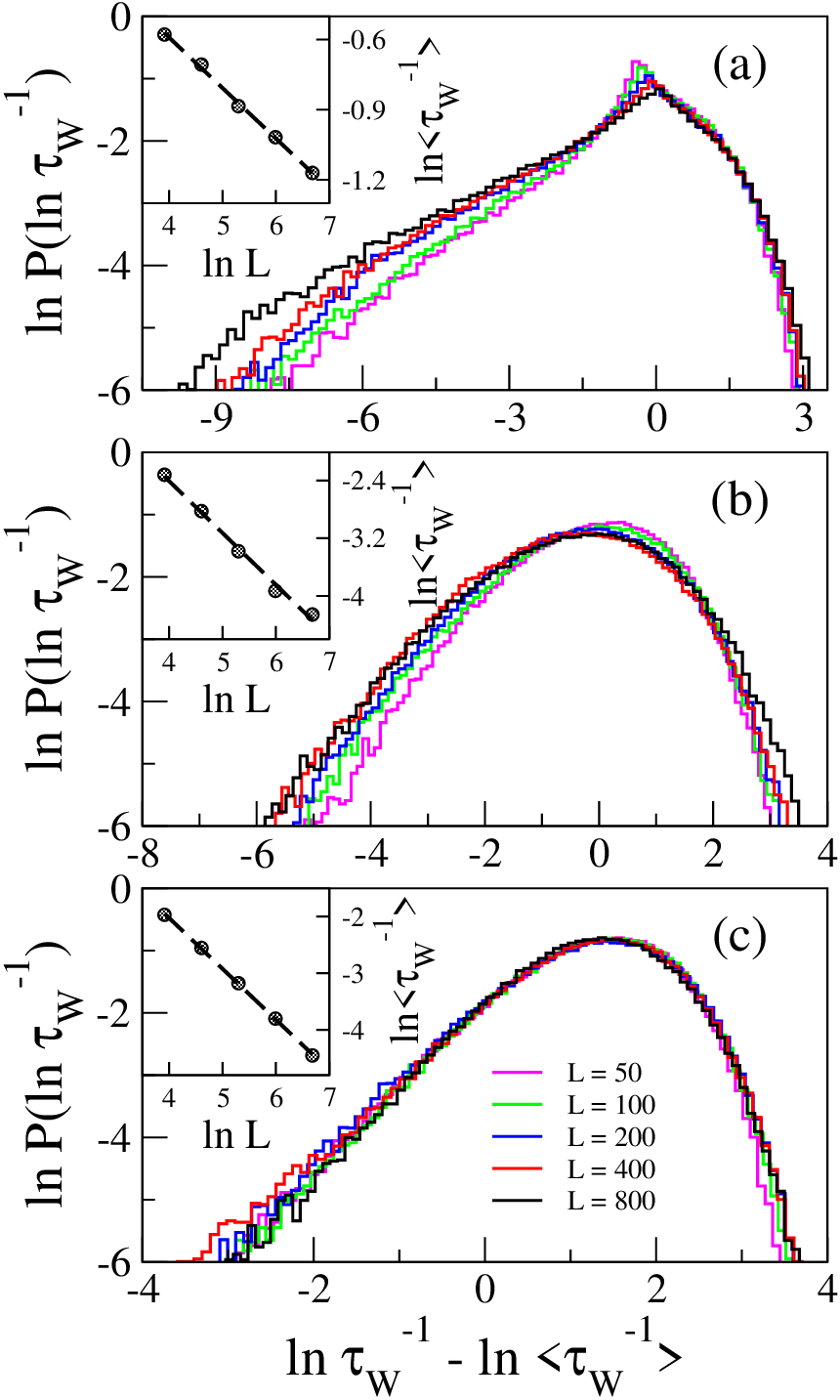}
\caption{$\ln {\cal P}(\ln \tau_W^{-1})$ as a function of $\ln \tau_W^{-1}$ scaled to $\ln \langle \tau_W^{-1} \rangle$ for a channel attached to a {\it representative} site (bulk) of the sample. (a) $b=0.1$, (b) $b=1$, and (c) $b=10$. In the insets the scaling $\langle \tau_W^{-1}\rangle \propto L^{-\alpha}$ is shown together with a linear fitting used to extract $\alpha$.}
\label{fig:fig2}
\end{center}
\end{figure}

In Fig. \ref{fig:fig2} we report the delay time distribution ${\cal P}(\ln \tau_W^{-1})$
for the case where the lead is attached to the bulk of the sample. Similar
distributions (not shown) are found for the case where the lead is attached to the
boundary of the sample.
We report results for various system sizes for bandwidths $b=0.1$, 1, and 10. The shape of
the distributions for a specific $b-$value changes with $L$, leading to the conclusion that various
moments of the distribution of inverse delay times scale in a different way with the system size. In the insets we concentrate on
the scaling of the first inverse moment $\langle \tau_W^{-1} \rangle\sim L^{-\alpha}$. A nice scaling
is observed. Using a least square fitting we can extract the scaling exponent $\alpha$.

\begin{figure}
\begin{center}
    \epsfxsize=8.4cm
    \leavevmode
    \epsffile{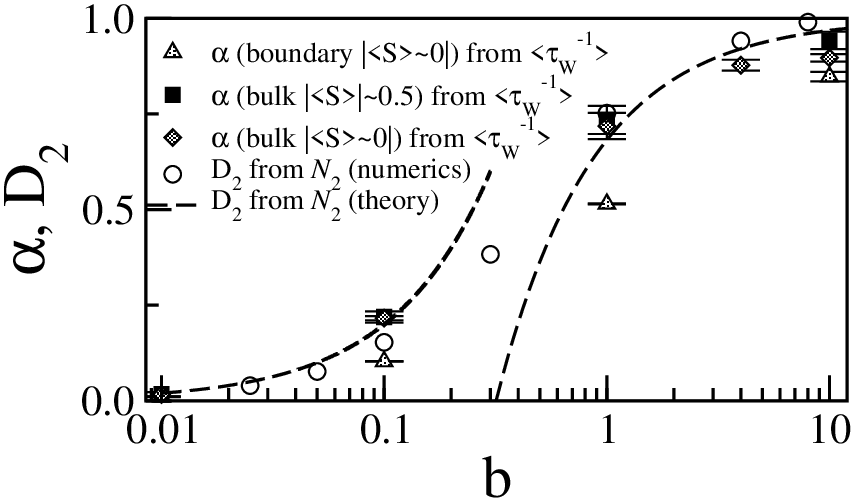}
\caption{Scaling exponent $\alpha$ and correlation dimension of eigenfunctions $D_2$ as a function of $b$.
$\alpha$ is extracted from the scaling $\langle \tau_W^{-1}\rangle \propto L^{-\alpha}$ for a channel attached to the bulk and to the boundary of the sample. $D_2$ calculated theoretically \cite{MFDQS96} and numerically \cite{V03} from ${\cal N}_2$ is also shown, see eq. (\ref{PN}).}
\label{fig:fig3}
\end{center}
\end{figure}

Our results are summarized in Fig.~\ref{fig:fig3} where we compare the scaling exponents obtained for the two
scattering setups (channel attached at the boundary and to the bulk) together with the theoretical
expectation (\ref{MIT}) which predicts that in the limit of $b\gg1$ the scaling exponent $\alpha=D_2$.
For completeness we include also the value of $D_2$ numerically extracted from the scaling analysis
of eigenvectors of the closed system \cite{V03}. We observe that Eq.~(\ref{MIT}) for the $\langle
\tau_W^{-1} \rangle$ valids quite well as long as the channel is connected to the bulk of the sample
where the ergodic hypothesis for the statistical properties of the eigenstates of the corresponding
closed system holds. Although the theoretical expression (\ref{MIT}) was derived under the assumption
of $b\gg 1$ we see that our numerical results for $b\ll 1$ can also be described by Eq.~(\ref{MIT}) relatively good.
Moreover, we find that the scaling relation (\ref{MIT}) persist for any value of $\langle S\rangle\neq 0$
in agreement with \cite{OF04}. On the contrary, for the case where a channel is attached to the boundary,
we observe strong deviations from Eq.~(\ref{MIT}). This is because the boundary is not a representative
position and the ergodicity hypothesis as far as the eigenfunctions is concerned is not any more applicable.

\begin{figure}
\begin{center}
    \epsfxsize=8.4cm
    \leavevmode
    \epsffile{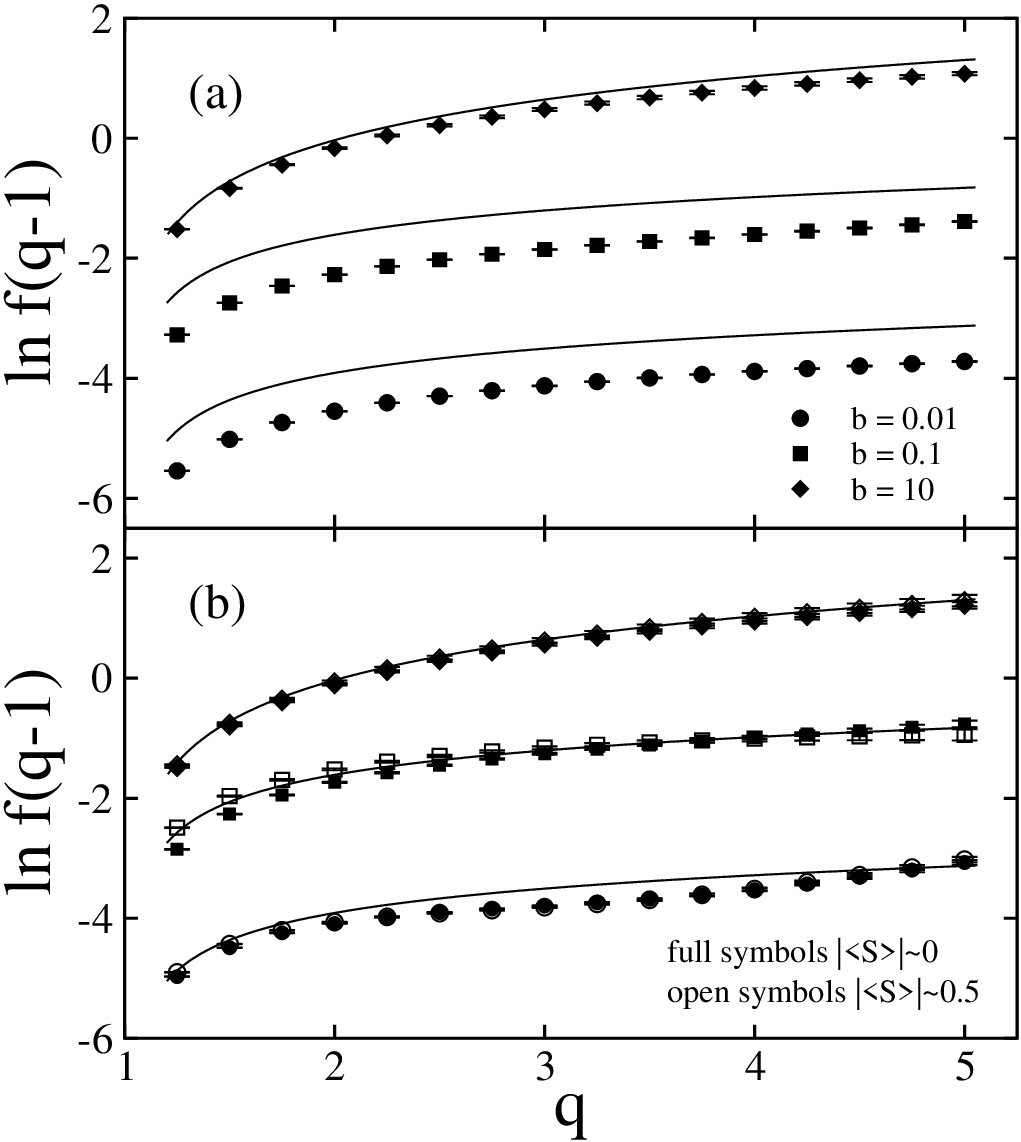}
\caption{$\ln f(q-1)$ as a function of $q$ for a channel attached (a) to the boundary and (b) to the bulk of the sample when $|\langle S\rangle|\sim 0$ (full symbols). In (b) we also show $\ln f(q-1)$ for $|\langle S\rangle|\sim 0.5$ (open symbols). The curves are the theoretical prediction of eq. (\ref{Dq}). $f(q-1)=(q-1)D_q$, see eq. (\ref{MIT}).}
\label{fig:fig4}
\end{center}
\end{figure}

Let us finally comment on the anomalous scaling behavior of the other moments
$\langle \tau_W^{-q} \rangle$. A simple homogeneous fractal would be completely characterized
by two of the moments and the respective $f(q)$ curve would be a straight line while a non-linearity
in $f(q)$ would signify a multifractal behavior. In Fig.~\ref{fig:fig4}(a) we report our
results for the case with a channel attached to the boundary. Again we see that the numerical data deviates
from the theoretical predictions (\ref{Dq}) for any value of $b$. Instead, the agreement is
quite good for the case where the channel is attached to the bulk of the sample (see Fig.~\ref{fig:fig4}b).

\section{Conclusions}

In summary we have studied the distribution of scattering phases ${\cal P}(\Phi)$ and delay times
for the power law banded matrix model at criticality. For small bandwidths $b\ll 1$ the distribution
of phases is non-uniform irrespective of the value of $|\langle S\rangle|$, while for large $b$
the theoretical prediction of \cite{OF04} applies.

We found that the multifractal nature of the eigenfunctions is reflected in the scaling of the inverse
moments of Wigner delay times $\langle \tau_W^{-q}\rangle$ with the system size $L$ provided that
the channel is attached to a {\it representative} position in the sample. We expect that these results
will provide a new method for evaluating $D_2$ (and in general $D_q$) in microwave and light wave experiments where $\tau_W$
can be extracted even in the presence of weak absorption. At the same time the appearance of the
anomalous scaling of $\langle \tau_W^{-q}\rangle $ can be used as a criterion for detecting MIT.

{\bf Acknowledgments}: We thank Y. Fyodorov and A. Ossipov for many discussions and suggestions related to the current study. M. Weiss is acknowledged for discussions on scattering.
This work was supported by the GIF, the German-Israeli Foundation for Research and Development.


\begin{thebibliography}{99}
\bibitem{note2} To our knowledge, the only experimental study of eigenfunction
fluctuations beyond the ballistic regime is in \cite{SS90}, where an analysis in
the weak localization regime was performed using the "microwave cavity" technique.
\end{thebibliography}
\end{document}